# Multi-echo AxCaliberSMT: Mapping effective axon radius from restricted diffusion and radius-dependent intra-axonal $T_2$

[Running title]

Joint diffusion-relaxometry for effective axon radius mapping

[Authors]

[1,2]Kwok-Shing Chan, PhD, kchan2@mgh.harvard.edu

[1,2]Yohan Jun, PhD, Jun, YJUN@mgh.harvard.edu

[1,2]Aneri Bhatt, abhatt7@mgh.harvard.edu

[1,2]Berkin Bilgic, PhD, BBILGIC@mgh.harvard.edu

[1,2]Susie Y. Huang, MD, PhD, susie.huang@mgh.harvard.edu

[1,2]Hong-Hsi Lee, MD, PhD, HLEE84@mgh.harvard.edu

[Affiliations]

[1]Department of Radiology, Athinoula A. Martinos Center for Biomedical Imaging, Massachusetts General Hospital, Charlestown, MA, United States

[2]Harvard Medical School, Boston, MA, United States

[Correspondence to]

Kwok-Shing Chan, PhD

Athinoula A. Martinos Center for Biomedical Imaging, Charlestown, MA, United States

149 13th St, Charlestown, MA, USA

Correspondence to Kwok-Shing Chan

# Abstract

**Purpose:** Diffusion-based axon radius *r* mapping preferentially detects larger axons since restricted-diffusion contrast scales with $r^4$, limiting sensitivity to thin axons that dominate human white matter. Surface relaxation provides complementary sensitivity through an intra-axonal $R_2$ proportional to 1/*r*. We combined these mechanisms to broaden the dynamic range of radius estimations.

**Methods:** Multi-echo AxCaliberSMT (ME-AxCaliberSMT) models spherical-mean diffusion-relaxation signals using cylinder-restricted diffusion and histologically calibrated surface relaxation. Simulations compared its accuracy and robustness with AxCaliberSMT and a diffusion-filtered $R_2$ model, which infer radius solely from restricted diffusion and intra-axonal $R_2$, respectively. Five healthy adults were scanned on Connectome 2.0 (500 mT/m), and four also on Cima.X (200 mT/m), using multiple b-values at two diffusion times, each paired with a different echo time.

**Results:** Diffusion and relaxation contrasts were comparable at a protocol-dependent transition radius (Connectome 2.0: 1.42 μm; Cima.X: 1.85 μm). Below this threshold, relaxation dominated, yielding effective radius estimates approximately $\langle r^2 \rangle / \langle r \rangle$ in most white matter. At SNR 40, ME-AxCaliberSMT achieved normalised bias within ±6% for radii of 0.1-3 μm, whereas AxCaliberSMT overestimated a radius of 0.25 μm by 342%. ME-AxCaliberSMT was less sensitive to tested calibration errors than the $R_2$ model. In vivo, ME-AxCaliberSMT bundle radii on Connectome 2.0 were 0.30-0.51 μm (AxCaliberSMT: 1.03-1.38 μm) and followed the histological pattern along the corpus callosum. ME-AxCaliberSMT bundle estimates correlated between scanners (mean Pearson correlation 0.73), unlike AxCaliberSMT (-0.23), although Cima.X values were systematically higher by 0.21 μm.

**Conclusion:** Combining diffusion and relaxation extends axon radius mapping from large to small axons. Translation to a commercial high-gradient scanner reproduced bundle-wise patterns, although absolute estimates remained protocol-dependent.

Correspondence to Kwok-Shing Chan

# 1. Introduction

Axon radius ($r$) is a key determinant of the conduction velocity of myelinated fibres[1] and is altered in neurological and neurodevelopmental disorders[2–7]. Histological studies show that most axons in human white matter have an inner radius below 1 μm, together with a small population of much larger axons[8,9]. A non-invasive measure of axon radius in this range would therefore be a valuable marker of white matter microstructure and its pathology.

Diffusion MRI infers effective axon radius from the restricted diffusion of water perpendicular to the axons. AxCaliber[10] and ActiveAx[11] model this signal directly. Spherical-mean approaches, including AxCaliberSMT[12], the power-law approach of Veraart et al.[13] and the single-encoding approach of Pizzolato et al.[14], remove the dependence on the fibre orientation distribution[15–17]. In the wide-pulse regime of human scanners[18,19], the intra-axonal radial diffusivity scales with $\sim r^4$, and radii below a resolution limit cannot be distinguished from zero at realistic SNR[20–22] (Figure 1a). Even with the 500 mT/m gradients of Connectome 2.0, the minimum detectable radius is about 1.25 μm[23], which is larger than most axons, and the limit is higher on commercial scanners with weaker gradient hardware. When a voxel contains a distribution of radii, the diffusion signal is also dominated by the largest axons[13,24] through the effective radius $r_D = \left(\frac{\langle r^6 \rangle}{\langle r^2 \rangle}\right)^{1/4}$.

Transverse relaxation provides a complementary route that is most sensitive where diffusion is least sensitive. In the fast-diffusion limit of surface relaxation[25], the membrane adds a term proportional to the surface-to-volume ratio, $2/r$, for a cylinder, so the intra-axonal relaxation rate becomes $R_{2,A} = R_2^C + k_{2,A}/r$, with the largest sensitivity for small axons (Figure 1b; see also Section 2.1). Simulations suggest that $T_2(=1/R_2)$ spectroscopy based on this effect could resolve axon diameters below 1 μm[26]. Intra-axonal $T_2$ can be separated from the other compartments with multi-echo, strongly diffusion-weighted or spherical-mean data[27–31], and Barakovic et al.[32] calibrated this model against histological radii in the corpus callosum. This $R_2$ model, however, uses only the echo-time dependence at a single high *b*-value. It ignores the

Correspondence to Kwok-Shing Chan

radius information in the diffusion signal, and residual extra-axonal signal biases the estimate.

In this work, we combine both mechanisms in a single model, called multi-echo AxCaliberSMT (ME-AxCaliberSMT), in which the axon radius determines both the restricted diffusion and the surface relaxation of the intra-axonal compartment, and each compartment has its own relaxation rate. We first derive the effective radius that the model estimates for a distribution of radii and then compare its accuracy and precision with those of AxCaliberSMT and the $R_2$ model in simulations and examine the bias caused by mis-specified $R_2^C$ and $k_{2,A}$. Finally, we apply the model to healthy volunteers scanned on Connectome 2.0 (500 mT/m) and on a commercial 3T system with 200 mT/m gradients, and assess how well bundle-wise radii agree between the two scanners.

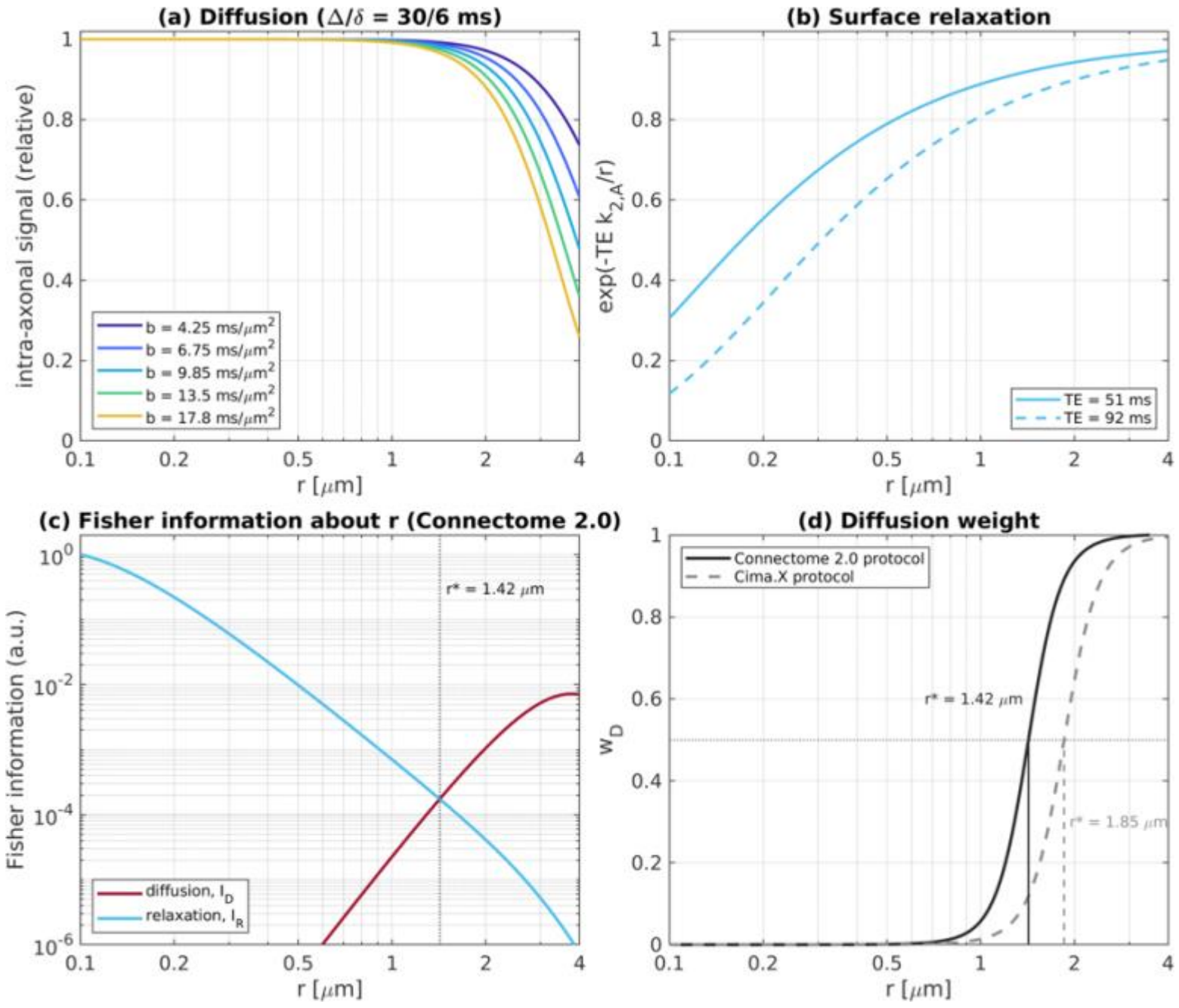

Figure 1: Radius contrast in ME-AxCaliberSMT. (a) Intra-axonal spherical-mean signal of a cylinder, relative to that of an infinitely thin cylinder, against radius for the long-diffusion-time shells of the Connectome 2.0 protocol ($\Delta/\delta$ = 30/6 ms); diffusion changes the signal appreciably only above about 1 μm. (b) Surface-relaxation factor $\exp(-TE \cdot k_{2,A}/r)$ at TE = 51 and 92 ms, which changes most for small radii. (c) Fisher information about $r$ carried by diffusion ($I_D$; Eq. [12]) and by relaxation ($I_R$; Eq. [12]) for the Connectome 2.0 protocol, with the other fitted parameters projected out (Supporting Information S1); the two are equal at the transition radius $r^*$ = 1.42 μm. (e) Diffusion weight $w_D$ (Eq. [12]) for the Connectome 2.0 and Cima.X protocols, with $r^*$ = 1.42 and 1.85 μm, respectively; radii below $r^*$ are determined mainly by relaxation and radii above it by diffusion.

# 2. Theory

## 2.1. Signal Model

ME-AxCaliberSMT describes the spherical-mean signal of each acquisition shell, defined by its *b*-value, diffusion time $\Delta$, gradient duration $\delta$ and echo time TE, as the sum of three non-exchanging compartments, namely intra-axonal (A), extra-axonal (E) and free water (CSF):

$$S(b,\Delta,\delta,TE) = S_0\{(1-f_{CSF})[f\,S_A + (1-f)\,S_E] + f_{CSF}\,S_{CSF}\} \quad Eq.[1]$$

where $S_0$ is the proton density-weighted signal intensity, $f$ is the intra-axonal fraction of the tissue (A+E) signal and $f_{CSF}$ is the free water fraction.

**Intra-axonal water:** The intra-axonal compartment is modelled as impermeable cylinders of radius $r$, whose spherical-mean signal is

$$S_A = \sqrt{\frac{\pi}{4b\left(D_{A,\parallel} - D_{A,\perp}\right)}}\; e^{-bD_{A,\perp}}\,\mathrm{erf}\left(\sqrt{b\left(D_{A,\parallel} - D_{A,\perp}\right)}\right)\; e^{-TE\,(R_2^C + k_{2,A}/r)} \quad Eq.[2]$$

where $D_{A,\perp}$ is the time-dependent radial diffusivity of a cylinder under pulsed field gradients in the Gaussian phase approximation[18]:

$$D_{A,\perp}(r,\Delta,\delta,D_0) = \frac{2}{D_0^2\,\delta^2\,(\Delta-\delta/3)}\sum_{m=1}^{\infty}\frac{2D_0\alpha_m^2\delta - 2 + 2e^{-D_0\alpha_m^2\delta} + 2e^{-D_0\alpha_m^2\Delta} - e^{-D_0\alpha_m^2(\Delta-\delta)} - e^{-D_0\alpha_m^2(\Delta+\delta)}}{\alpha_m^6\,(r^2\alpha_m^2 - 1)} \quad Eq.[3]$$

with $D_0$ the intrinsic axoplasmic diffusivity and $\alpha_m r$ the roots of the derivative of the first-order Bessel function of the first kind. In the wide-pulse limit[19] ($\delta \gg r^2/D_0$), $D_{A,\perp}$ reduces to $7/48 \cdot r^4/(D_0\delta(\Delta-\delta/3))$.

The intra-axonal relaxation rate follows the surface-relaxation model of Barakovic et al.[32], with an intrinsic axoplasm rate $R_2^C$ and a surface term $k_{2,A}/r$, where $k_{2,A} = 2\rho_2$ and $\rho_2$ is the T2 surface relaxivity. In the wide-pulse regime, $D_{A,\perp}$ scales with $\sim r^4$, while the radius-dependent part of $R_{2,A}$ scales with $\sim 1/r$, so the diffusion contrast is sensitive to large radii and the relaxation contrast to small radii.

**Extra-cellular water:** The extra-axonal compartment is modelled as an axially symmetric tensor (zeppelin) with parallel and radial diffusivities $D_{E,\parallel}$ and $D_{E,\perp}$ and relaxation rate $R_{2,E}$:

$$S_E = \sqrt{\frac{\pi}{4b\left(D_{E,\parallel} - D_{E,\perp}\right)}}\ e^{-bD_{E,\perp}} \operatorname{erf}\left(\sqrt{b\left(D_{E,\parallel} - D_{E,\perp}\right)}\right)\ e^{-TE \cdot R_{2,E}} Eq. [4]$$

**Free water:** The free water compartment is modelled as an isotropic Gaussian compartment:

$$S_{CSF} = e^{-bD_{CSF}}\ e^{-TE \cdot R_{2,CSF}} Eq. [5]$$

Because $T_2$ relaxation is modelled explicitly, $f$ and $f_{CSF}$ are proton-density fractions, unlike those of AxCaliberSMT[12]. Without the relaxation terms, Eq. [1] reduces to AxCaliberSMT. The $R_2$ model[32] keeps only the TE dependence of Eq. [2], $S(TE) = \hat{S}_A e^{-TE\left(R_2^C + k_{2,A}/r\right)}$ with $\hat{S}_A$ contains the diffusion weighted contrast, and assumes that the extra-axonal and free water signals are highly suppressed at high *b*-values. Table 1 summarises the model parameters.

Table 1: ME-AxCaliberSMT model parameter.

| Parameter | Meaning | Estimated or fixed | Range or value |
|---|---|---|---|
| $r$ | (Effective) Axon radius | Estimated | 0-5 μm |
| $f$ | intra-axonal fraction of tissue signal | Estimated | 0-1 |
| $f_{CSF}$ | CSF signal fraction | Estimated | 0-1 |
| $D_{E,\perp}$ | Extra-axonal radial diffusivity | Estimated | 0-1.7 $\mu m^2/ms$ |
| $R_{2,E}$ | Extra-axonal $R_2$ | Estimated | 1-50 $s^{-1}$ |
| $R_2^C$ | Intrinsic axoplasm $R_2$ | Fixed | 7.88 $s^{-1}$ |
| $k_{2,A}$ | $T_2$ surface relaxivity | Fixed | 2.32 μm/s |
| $D_0 = D_{A,\parallel} = D_{E,\parallel}$ | Intrinsic, intra-axonal and extra-axonal parallel diffusivity | Fixed | 1.7 $\mu m^2/ms$ |
| $D_{CSF}$ | Free water diffusivity | Fixed | 3 $\mu m^2/ms$ |
| $R_{2,CSF}$ | Free water $R_2$ | Fixed | 0.5 $s^{-1}$ |

## 2.2. Effective radius with a distribution of axon radii

To interpret the estimated radius, we consider how the intra-axonal signal (Eq. [2]) depends on $r$ through restricted diffusion and through surface relaxation, two mechanisms with opposite radius dependence. For weak radial attenuation ($bD_{A,\perp} \ll 1$), the sensitivity through diffusion is

$$\left|\frac{\partial S}{\partial r}\right|_D \propto \frac{7}{12} \frac{b\, r^3}{D_0\, \delta\, (\Delta - \delta/3)} Eq.[6]$$

which vanishes for small radii, whereas the sensitivity through relaxation

$$\left|\frac{\partial S}{\partial r}\right|_R \propto TE\, \frac{k_{2,A}}{r^2} \propto r^{-2} Eq.[7]$$

is largest for small radii. Since $f$ and $R_{2,E}$ are estimated, only the change in signal between echo times carries information on $r$ through relaxation, and the echo spacing ΔTE takes the place of TE. The ratio of the two sensitivities is then

$$\frac{|\partial S/\partial r|_D}{|\partial S/\partial r|_R} = \frac{7}{12} \frac{b\, r^5}{\Delta TE\;\, k_{2,A}\, D_0\, \delta\, (\Delta - \delta/3)} Eq.[8]$$

and the two sensitivities are equal at the transition radius

$$r^* = \left(\frac{12}{7} \frac{\Delta TE\;\, k_{2,A}\, D_0\, \delta\, (\Delta - \delta/3)}{b}\right)^{1/5} \quad Eq.[9]$$

is a function of acquisition settings and surface relaxivity. The Fisher information scales with the squared sensitivity, so the information from diffusion relative to that from relaxation scales as $(r/r^*)^{10}$. Radii well below $r^*$ are therefore determined by relaxation, and radii well above it by diffusion. Because of the fifth root, $r^*$ depends only weakly on the protocol; doubling b lowers it by 13%. For the strongest diffusion weighting on Connectome 2.0 ($b$ = 17.8 ms/μm$^2$, Δ = 30 ms, $\delta$ = 6 ms, ΔTE = 41 ms), Eq. [9] gives $r^*$ ≈ 1.2 μm.

For voxels containing a range of axons with radii $r_i$, each axon contributes to the intra-axonal signal in proportion to its cross-sectional area, $r_i^2$, and each contrast reflects a different effective radius. Expanding the relaxation term to first order in $TE \cdot k_{2,A}/r$, which is valid when $\Delta TE \cdot k_{2,A}$ (about 0.1 μm) is small compared with the radii, gives the relaxation effective radius:

Correspondence to Kwok-Shing Chan

$$r_R = \frac{\langle r^2 \rangle}{\langle r \rangle} \quad Eq. [10]$$

For weak radial attenuation, the diffusion signal is determined by the area-weighted mean of $r^4$, which defines the diffusion effective radius[13,24]:

$$r_D \approx \left( \frac{\langle r^6 \rangle}{\langle r^2 \rangle} \right)^{1/4} \quad Eq. [11]$$

Because higher moments weight large axons more strongly, $r_D \geq r_R$, with equality only for a single radius.

ME-AxCaliberSMT estimates a single effective radius $\hat{r}$ from data that reflect both effective radii. Linearising the least-squares problem around the solution (see Supporting Information S1), with the nuisance parameters ($f$, $f_{CSF}$, $D_{E,\perp}$, $R_{2,E}$) projected out, yields

$$\hat{r} \approx w_D \, r_D + (1 - w_D) \, r_R \quad Eq. [12]$$

where $w_D = I_D/(I_D + I_R)$, and $I_D = \sum_j \left| \frac{\partial S_j}{\partial r} \right|_D^2 / \sigma_n^2$ and $I_R = \sum_j \left| \frac{\partial S_j}{\partial r} \right|_R^2 / \sigma_n^2$ are the Fisher information on $r$ from the diffusion and relaxation parts of the signal ($\sigma_n$ is the standard deviation of noise; Figure 1c). From Eq. [8], $w_D$ rises from 0 to 1 over a narrow range of radii around $r^*$: below it, ME-AxCaliberSMT reports $r_R$, which is weighted much less towards large axons, and above it, the estimate approaches the AxCaliberSMT radius $r_D$. For a single radius, $\hat{r} = r$ regardless of $w_D$.

# 3. Methods

## 3.1. Simulations

### 3.1.1. Numerical Simulation #1: Radius estimation across a wide range of radii

To isolate the effect of axon size, all axons in a voxel had the same radius ($\sigma(r) = 0$), and we asked whether a second TE allows ME-AxCaliberSMT to recover thin axons below the resolution limit of AxCaliberSMT[21,22].

Signals were simulated with Eqs. [1-5] for the Connectome 2.0 protocol (Table 2; 902 volumes at TE = 51 and 92 ms); for AxCaliberSMT, all volumes were simulated at TE = 51 ms, which gives it an SNR advantage. In each of 10,000 voxels, the parameters were drawn from uniform distributions: $r \in [0.1, 3]$ µm, $f \in [0.1, 0.8]$, $f_{CSF} \in [0, 0.1]$, $R_{2,E} \in [17, 50]$ s$^{-1}$ and Watson dispersion $p_2 \in [0.1, 1]$. $D_{E,\perp}$ was set by first-order tortuosity, $D_{E,\perp} = D_{E,\parallel}(1 - f)$[33], but was free in the fit, and the remaining parameters were fixed as in Table 1. The intra-axonal signal was computed for every gradient direction with the wide-pulse cylinder model[18] and a dispersed fibre orientation distribution (spherical harmonics up to l = 8). Gaussian noise was added at SNR = 40, defined at *b* = 0 and TE = 51 ms, with the same noise realisation for both models.

Parameters were estimated from the spherical mean of each shell with the sampler described in Section 3.4, with $R_2^C$ and $k_{2,A}$ fixed to their ground-truth values and the posterior median as the estimate. The simulation was repeated five times with new parameters and noise.

Accuracy was quantified for every fitted parameter $\theta$ as the normalised bias, $(\hat{\theta} - \theta_{GT})/\theta_{GT}$, and precision as the standard deviation (SD) of $\hat{\theta} - \theta_{GT}$, both in ten equal-width bins of the ground truth (bin width 0.29 µm for $r$) and given as mean ± SD over the five repetitions; for $f_{CSF}$, the bins started at 0.01 because the normalised bias diverges near zero. The AxCaliberSMT fractions were also compared with the ground-

Correspondence to Kwok-Shing Chan

truth fractions weighted by the $T_2$ decay of each compartment at TE = 51 ms, and the overall error was summarised by the root mean square error (RMSE).

### 3.1.2. Numerical Simulation #2: Sensitivity to mis-specified $R_2^C$ and surface relaxivity $k_{2,A}$

The constants $R_2^C$ and $k_{2,A}$ are fixed to the calibration of Barakovic et al.[32] but may vary between subjects and tissues. We therefore tested how much each model degrades when they are mis-specified.

Signals were generated as in Section 3.1.1 for six ground-truth conditions: no mismatch, $R_2^C$ and $k_{2,A}$ 10% higher or lower than assumed, and both constants drawn independently for each voxel within ±10%. The fit always assumed $R_2^C$ = 1/127 ms$^{-1}$ and $k_{2,A}$ = 2.32 μm/s. For the $R_2$ model, the same number of volumes was simulated at *b* = 6 ms/μm$^2$ ($\Delta/\delta$ = 13/6 ms), distributed over TE = 51, 92, 118 and 150 ms (225-227 directions per TE), and the model was fitted to the spherical mean at all four TEs with the same sampler. Tissue parameters, noise and all other settings followed Section 3.1.1.

### 3.1.3. Validation of effective radius

We test Eq. [12] at negligible noise. Within each voxel, radii followed a gamma distribution with a mean of 0.25-2 μm (in steps of 0.25 μm) and $CoV(r) = \sigma(r)/\langle r \rangle$ of 0, 0.25 or 0.5, truncated at 5 μm (24 conditions), and the intra-axonal signal was the area-weighted sum of Eq. [2] over all radii. The other parameters were fixed ($f$ = 0.7, $f_{CSF}$ = 0.05, $D_{E,\perp}$ = 0.51 μm$^2$/ms and $R_{2,E}$ = 25 s$^{-1}$, with $R_2^C$ and $k_{2,A}$ as in Table 1). For each condition, 50 voxels were simulated at SNR = 1000 with the protocols of Section 3.1.1-2 and fitted as in Section 3.1.1.

The median fitted radius was compared with $r_R$ from Eq. [10], with $r_D$ from Eq. [11]. We also computed $r_D$ numerically as the single effective radius whose cylinder signal

Correspondence to Kwok-Shing Chan

best fits the diffusion signal of the mixture, and with their combination in Eq. [12], with $w_D$ evaluated at the fitted radius by fixed-point iteration (see Supporting Figure S1). Agreement was summarised as the RMS difference over the 16 conditions with $CoV(r) > 0$. Supporting Information S1 gives the details, including $w_D(r)$ and $r^*$ (where $w_D$ = 0.5) for the in vivo protocols. AxCaliberSMT was compared with $r_D$ and the $R_2$ model with $r_R$.

## 3.2. In vivo Imaging

Five healthy volunteers (three female, two male; age 22-45 years) were scanned on the 3T Connectome 2.0 scanner[34,35] (MAGNETOM Connectom.X, Siemens, Germany) with a 72-channel head coil[36], and four of them were also scanned (mean ± SD time between the two sessions: 201 ± 49 days) on a 3T MAGNETOM Cima.X (Siemens, Germany) with a 64-channel head/neck coil. All participants gave written informed consent, and the study was approved by the institutional review board. Each session included:

- a 1 mm isotropic $T_1$-weighted MPRAGE (TR/TI/TE = 2500/1100/3.3 ms, flip angle = 8°, GRAPPA factor 2; acquisition time = 5.5 min)
- dMRI acquired with a monopolar pulsed-gradient spin-echo sequence and a single-shot two-dimensional echo-planar readout (2 mm isotropic). Each protocol used two diffusion times and two echo times (Table 2). Non-diffusion-weighted ($b$ = 0) images were interspersed every 16 diffusion-weighted images, and $b$ = 0 images with reversed phase encoding were acquired at each TE for distortion correction. Total acquisition time = 68 min (C2) and 66 min (Cima.X).

For the $R_2$ model, four participants also underwent a multi-echo acquisition on Connectome 2.0 with the same sequence (2.5 mm isotropic), acquired as separate scans at TE = 51, 92, 118 and 150 ms (TR = 5100 ms) together with a $b$ = 0 image with reversed phase encoding, with an acquisition time of 24 min.

Table 2: Diffusion protocols for in vivo experiment. *TE = 150 ms was acquired but not used due to low SNR and image artefact.

| | **ME-AxCaliberSMT** | **$R_2$ model** |
|---|---|---|
| **Diffusion time, Δ (ms)** | Connectome 2.0<br>13, 30<br>Cima.X<br>22.5, 35 | 13 |
| **Diffusion gradient duration, $\delta$ (ms)** | Connectome 2.0<br>6<br>Cima.X<br>15 | |
| **#Diffusion gradient direction per shells** | 16 ($b$ < 2.4 ms/μm$^2$)<br>32 ($b \geq$ 2.4 ms/μm$^2$) | 64 |
| ***b* (ms/μm$^2$)** | Connectome 2.0<br>$\Delta_{13}$: 0.05, 0.35, 0.8, 1.5, 2.4, 3.45, 4.75, 6<br>$\Delta_{30}$: 0.2, 0.95, 2.3, 4.25, 6.75, 9.85, 13.5, 17.8<br>Cima.X<br>$\Delta_{22.5}$: 0.05, 0.35, 0.8, 1.5, 2.4, 3.45, 5<br>$\Delta_{35}$: 0.6, 1.35, 2.6, 4.1, 6 | 6 |
| **TR/TE** | Connectome 2.0<br>4500/[51, 92]<br>Cima.X<br>6200/[64, 92] | 5100/[51, 92, 118, 150*] |
| **SMS/$R_{GRAPPA}$/Partial Fourier** | 2/2/0.75 | |
| **Resolution (mm)** | 2 isotropic | 2.5 isotropic |
| **Acquisition time (min)** | Connectome 2.0<br>68<br>Cima.X<br>66 | 24 |

## 3.3. Data Processing

All acquisitions, including the multi-echo acquisition for the $R_2$ model, were denoised with MP-PCA[37] and an iterative correction of the Rician noise floor (in-house implementation, 10 iterations)[38]. Gibbs ringing was removed from the *b* = 0 images[39] with MRtrix3[40]. Susceptibility-induced distortions were estimated with FSL topup[41], and eddy-current distortions and head motion were corrected with FSL eddy, including slice-to-volume correction and outlier replacement[42–44]. Gradient non-linearity distortions were then corrected with the scanner-specific coefficients[45,46].

Brain masks were obtained with SynthStrip[47] and tissue labels with SynthSeg[48,49]. White matter bundles were segmented with TractSeg[50] on the Connectome 2.0 data, using fibre orientation peaks from multi-shell multi-tissue constrained spherical deconvolution[51] of the short-Δ acquisitions at TE = 51 ms ($b \leq 6$ ms/μm$^2$). Cima.X maps were transformed into Connectome 2.0 diffusion space with ANTs[52] and linear interpolation.

## 3.4. Parameter estimation

For each shell (a unique combination of $b$, $\Delta$, $\delta$ and TE), the spherical mean was computed as the l = 0 spherical harmonic coefficient and normalised to the $b$ = 0 signal of the voxel at the shortest TE. For ME-AxCaliberSMT, five parameters were estimated ($f$, $f_{CSF}$, $D_{E,\perp}$, $r$ and $R_{2,E}$), and the others were fixed (Table 1), with $R_2^C$ and $k_{2,A}$ set to the histological calibration[32]. We used an adaptive Metropolis-Hastings sampler implemented in GACELLE[53], with a Gaussian likelihood, the noise marginalised under a Jeffreys prior[54], and parameters constrained to the bounds in Table 1. Two over-dispersed chains of 40,000 iterations (20,000 burn-in, thinning factor 40) gave 1,000 posterior samples per voxel, and the posterior median was used as the estimate. A whole-brain fit took 9-11 min.

AxCaliberSMT was fitted with the same sampler and settings, using the model without relaxation terms and only the short-TE acquisitions (451 volumes at TE = 51 ms on Connectome 2.0 and 318 at TE = 64 ms on Cima.X). The $R_2$ model was fitted to the spherical mean at $b$ = 6 ms/μm$^2$ of the multi-echo acquisition. Only the first three TEs (51, 92 and 118 ms) were used, because the SNR at 150 ms was too low and quality was severely degraded by imaging artefacts. $R_2$ was estimated by log-linear least squares and converted to r with the same $R_2^C$ and $k_{2,A}$, and the maps were registered to the main acquisition with a rigid-body transformation.

## 3.4. Statistical Analysis

White matter was defined by the SynthSeg labels of cerebral and cerebellar white matter, brainstem and ventral diencephalon. For the comparison between scanners, each TractSeg bundle was intersected with this mask and with the valid-voxel masks of both sessions in Connectome 2.0 space; after excluding bundles with fewer than 20 voxels, the fornix (CSF partial volume) and the union of the corpus callosum segments, 69 bundles per participant remained. The bundle value was the median of the voxel-wise posterior medians. Agreement was assessed with the Pearson and Spearman correlations across bundles within each participant, averaged over participants (also used to compare the two models on each scanner), Lin's concordance correlation coefficient (CCC)[55] and the Bland-Altman bias and 95% limits of agreement[56]. Because bundles overlap, their values are not independent, and we report no *p*-values. For the three-model comparison on Connectome 2.0, bundle medians were averaged over the four participants with the multi-echo acquisition to compute the correlation between models across the 69 bundles (each TractSeg bundle used directly); the SD across participants was computed from the individual values.

Correspondence to Kwok-Shing Chan

# 4. Results

## 4.1 Accuracy and precision across radii

Across radii of 0.1-3 μm at an SNR of 40, the normalised bias of ME-AxCaliberSMT stayed within ±6% in every radius bin (Figure 2a). AxCaliberSMT, with the same number of volumes, lost sensitivity to small radii, with an overestimation of 342% at 0.25 μm and 57% at 0.54 μm on normalised bias, and underestimated radii of 1-2 μm by 9-24%. The precision of ME-AxCaliberSMT, measured as the SD of fitting bias, was best for small and large radii (0.05 μm below 0.4 μm and 0.11 μm above 2.7 μm) and poorest at 1-2.4 μm (0.19-0.22 μm), where the relaxation contrast has weakened but the diffusion contrast is not yet strong (Section 2.2). Over all radii, the RMSE was 0.18 μm for ME-AxCaliberSMT and 0.41 μm for AxCaliberSMT, and the two models agreed to within 2% above 2.5 μm.

ME-AxCaliberSMT estimated the proton-density fractions with a normalised bias of +2% to +7% for $f$ and within ±4% for $f_{CSF}$ (Figure 2b-c). AxCaliberSMT overestimated $f$ by 18-160% and $f_{CSF}$ by 132-215%, more so at small fractions, consistent with the ground-truth fractions weighted by the $T_2$ decay of each compartment at TE = 51 ms. Both models overestimated the smallest extra-axonal radial diffusivities (by 30% and 37% at 0.40 μm$^2$/ms), where the slowly decaying extra-axonal signal is difficult to separate from the intra-axonal signal, but were within 5% above 1 μm$^2$/ms (Figure 2d). The normalised bias of $R_{2,E}$ was within +3% below 35 s$^{-1}$ and −21% at 48 s$^{-1}$, close to the prior boundary (Figure 2e).

Correspondence to Kwok-Shing Chan

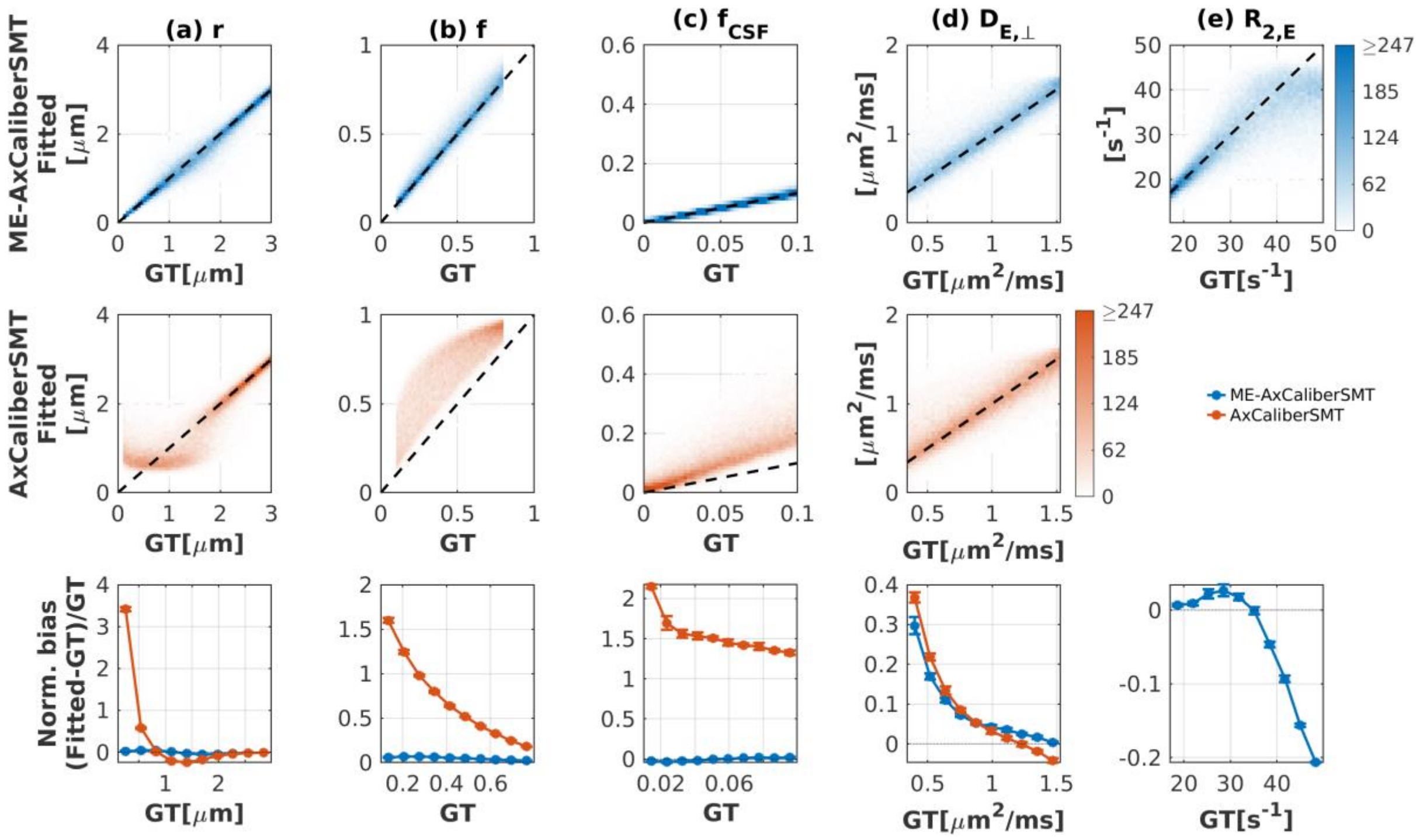

Figure 2: Accuracy and precision across radii for ME-AxCaliberSMT (top, in blue) and AxCaliberSMT (middle, in red). Density scatter plots of the fitted against the ground-truth parameters for (a) $r$, (b) $f$, (c) $f_{CSF}$, (d) $D_{E,\perp}$ and (e) $R_{2,E}$, pooled over five repetitions of 10,000 voxels at an SNR of 40. Both models used the same 902 diffusion encodings, at TE = 51 and 92 ms for ME-AxCaliberSMT and all at TE = 51 ms for AxCaliberSMT. Dashed lines indicate identity. Bottom row: normalised bias (fitted − ground truth)/ground truth, in ten bins of the ground truth (mean ± SD over repetitions).

## 4.2 Sensitivity to $R_2^C$ and $k_{2,A}$

With correctly specified constants, the $R_2$ model weakly overestimated $r$ by 3-13%, and the SD of its error increased from 0.05 μm below 0.4 μm to 0.68 μm above 2.7 μm; the normalised bias of ME-AxCaliberSMT stayed within ±6% (Figure 3a).

A 10% error in $k_{2,A}$ rescaled $r$ where relaxation dominated (Figure 3b-c). When $k_{2,A}$ was 10% lower than assumed, both models overestimated $r$ by 14-15% below 0.4 μm; for ME-AxCaliberSMT, this error fell to within ±4% above 1.3 μm, whereas the $R_2$ model remained biased by 9-25% at all radii. When $k_{2,A}$ was 10% higher, ME-AxCaliberSMT underestimated $r$ by 1-8%, and the $R_2$ model by up to 6%. Yet, the precision of $R_2$ model remained lower compared to ME-AxCaliberSMT.

In contrast, a 10% error in $R_2^C$ had a larger effect, because it shifts the fitted $1/r$ by $\Delta R_2^C/k_{2,A}$ ≈ 0.34 μm$^{-1}$, an error that grows with $r$ (Figure 3d-e). With $R_2^C$ 10% lower than assumed, the R2 model overestimated $r$ by up to 109% at 1.55-1.84 μm, whereas the normalised bias of ME-AxCaliberSMT peaked at +35% at 0.68-0.97 μm and was within +1% above 2.1 μm. With $R_2^C$ 10% higher, the R2 model underestimated $r$ by up to 39% above 2.7 μm, and ME-AxCaliberSMT by up to 26% at 1.26-1.55 μm and by at most 4% above 2.4 μm. When both constants varied randomly by ±10%, the normalised bias of ME-AxCaliberSMT stayed within ±7%, with lower precision only at 1-2.1 μm (SD: 0.24-0.27 vs 0.19-0.21 μm), whereas the R2 model overestimated $r$ by up to 22% (Figure 3f).

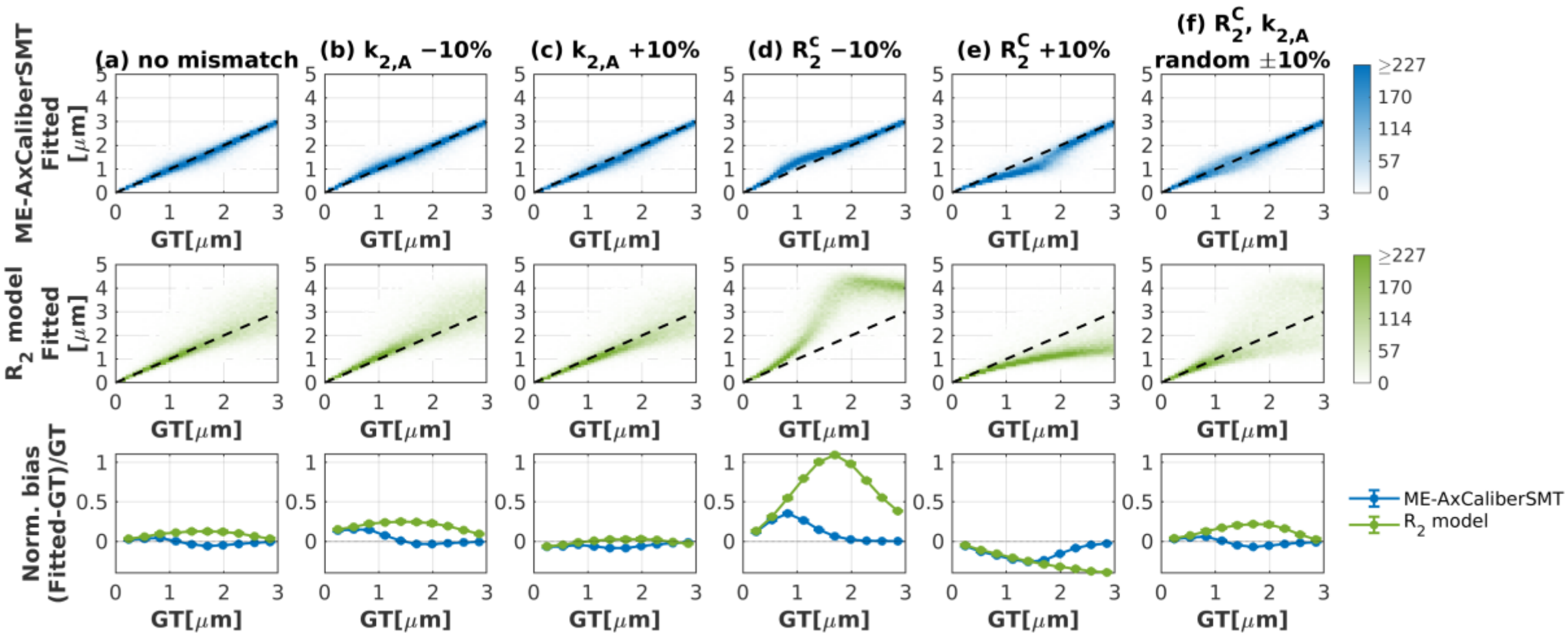

Figure 3: Sensitivity of the radius estimate to mis-specified $R_2^C$ and $k_{2,A}$. (a) No mismatch, with the ground-truth constants equal to the values assumed in the fit ($R_2^C$ = 1/127 ms$^{-1}$, $k_{2,A}$ = 2.32 μm/s); (b, c) ground-truth $k_{2,A}$ 10% lower and higher; (d, e) ground-truth $R_2^C$ 10% lower and higher; and (f) both constants drawn randomly within ±10% for each voxel. Density scatter plots of the fitted against the true radius for ME-AxCaliberSMT (top) and the R2 model (middle) at an SNR of 40, pooled over five repetitions; dashed lines indicate identity. Bottom row: normalised bias of r in ten bins of the true radius (mean ± SD over repetitions).

## 4.3 Effective radius of ME-AxCaliberSMT

The diffusion weight $w_D$ increased from 0 to 1 around $r^*$ = 1.42 μm on Connectome 2.0 ($w_D$ = 0.1-0.9 between 1.08 and 1.88 μm) and $r^*$ = 1.85 μm on Cima.X (1.39-2.48 μm; Figure 1d), compared with 1.21 and 1.70 μm from Eq. [9] for the strongest diffusion weighting. At $r$ = 1 µm, $w_D$ was 0.06 on Connectome 2.0 and 0.01 on Cima.X, so radii typical of white matter are determined mainly by relaxation on both scanners.

For a single radius at an SNR of 1000 (Figure 4a), ME-AxCaliberSMT recovered $r$ to within 0.003 µm between 0.25 and 2 µm, whereas AxCaliberSMT estimated true radii of 0.25, 0.5 and 0.75 µm as 0.34, 0.41 and 0.65 µm. The $R_2$ model underestimated $r$ by 0.04 µm at 1 µm and by 0.16 µm at 2 µm because of residual extra-axonal signal at *b* = 6 ms/µm$^2$.

For distributions of radii (Figure 4b-c), ME-AxCaliberSMT followed $r_R$ at small mean radii, with estimates 0.01-0.06 µm above $r_R$ for $\langle r\rangle$ ≤ 0.75 µm, and converged to AxCaliberSMT at large ones, differing from it by at most 0.03 µm for $\langle r\rangle$ ≥ 1.5 µm; the switch took place as the estimate crossed the transition range of $w_D$ (grey band in Figure 4). Over the 16 conditions with CoV > 0, ME-AxCaliberSMT followed Eq. [12] with the numerical $r_D$ to within an RMS deviation of 0.022 µm, against 0.30 µm for $r_R$ alone and 0.24 µm for the $r_D$ of Eq. [11] alone (Supporting Figure S1). The $R_2$ model followed $r_R$ to within 0.04 µm for $\langle r\rangle$ ≤ 0.75 µm but fell increasingly below it at larger mean radii, by up to 23% at CoV = 0.5 and $\langle r\rangle$ = 2 µm (1.92 vs 2.49 µm), because it neglects intra-axonal diffusion: weighting Eq. [10] by the diffusion attenuation at *b* = 6 ms/µm² lowers $r_R$ to 2.14 µm.

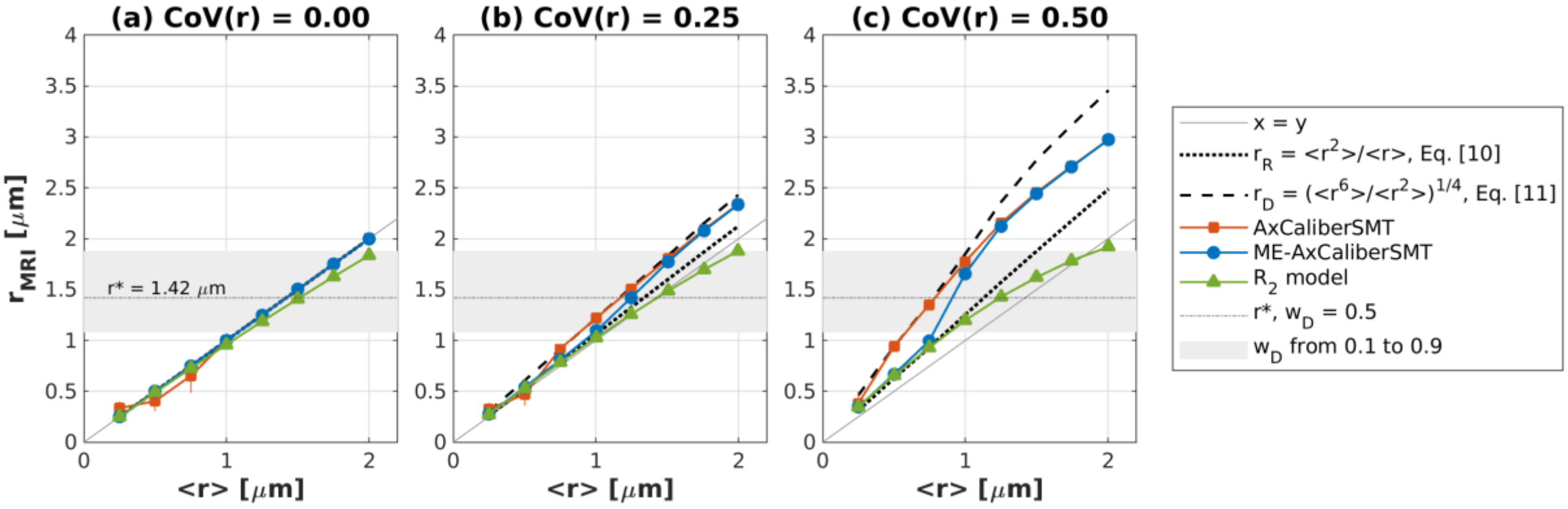

Figure 4: Effective radius for distributions of axon radii. Radius estimated by ME-AxCaliberSMT, AxCaliberSMT and the $R_2$ model against the mean radius $\langle r\rangle$ of gamma distributions with (a) CoV($r$) = 0, (b) 0.25 and (c) 0.5 at an SNR of 1000 (median over 50 voxels of the posterior median; bars: interquartile range across voxels). Reference curves show identity, the relaxation effective radius $r_R = \langle r^2\rangle/\langle r\rangle$ (Eq. [10]) and the diffusion effective radius $r_D = (\langle r^6\rangle/\langle r^2\rangle)^{1/4}$ (Eq. [11]). The dash-dotted line

marks the transition radius $r^*$ = 1.42 μm of the Connectome 2.0 protocol, and the grey band the range of estimated radii over which $w_D$ increases from 0.1 to 0.9 (1.08-1.88 μm).

## 4.4 In vivo experiments

In the radius maps of one example participant on Connectome 2.0, all three models showed larger radii along the corticospinal tract (CST) than in the surrounding white matter, with AxCaliberSMT highest and the $R_2$ model lowest (blue arrows, Figure 5). ME-AxCaliberSMT followed AxCaliberSMT along the tract and was closer to the $R_2$ model in frontal white matter, where small axons predominate (green arrows). In the midsagittal corpus callosum (CC), it reproduced the low-high-low-high pattern from genu to splenium known from histology[9] (black arrows), which the $R_2$ model showed over a narrower range and AxCaliberSMT less clearly.

For the other parameters, both models gave the expected contrast between white and grey matter in $f$ and similar $D_{E,\perp}$ maps (Figure 6). The AxCaliberSMT fractions, including $f_{CSF}$, were higher throughout white matter, whereas $f_{CSF}$ of ME-AxCaliberSMT was close to zero outside the ventricles and sulci, as expected from the $T_2$ weighting in Figure 2b-c. $R_{2,E}$ was lower in deep white matter than near the cortex and the ventricles.

In the four participants with both the ME-AxCaliberSMT and $R_2$ model acquisitions, AxCaliberSMT gave the largest radius in every bundle (1.03-1.38 μm) and the $R_2$ model the smallest (0.22-0.42 μm; Figure 7). ME-AxCaliberSMT lay in between, close to the $R_2$ model (0.30-0.51 μm). Across the 69 TractSeg bundles, the ME-AxCaliberSMT bundle values correlated strongly with those of AxCaliberSMT (Pearson correlation = 0.80) and more weakly with those of the $R_2$ model (0.48), the same correlation as between AxCaliberSMT and the $R_2$ model (0.48). AxCaliberSMT had the smallest contrast between bundles (CoV = 0.10, against 0.16 for both ME-AxCaliberSMT and the $R_2$ model) and the largest variability between participants (SD up to 0.26 μm, against at most 0.07 and 0.04 μm).

Correspondence to Kwok-Shing Chan

Along the CC, ME-AxCaliberSMT radii followed the low-high-low-high pattern (0.30, 0.35, 0.33, 0.41, 0.46, 0.42 and 0.51 µm from rostrum to splenium), whereas the $R_2$ model increased from 0.22 µm in the rostrum to 0.39 µm in the posterior midbody but stayed low in the isthmus and splenium (0.30 and 0.32 µm). Outside the CC, ME-AxCaliberSMT radii were largest in the CST (0.49 µm) and smallest in the cingulum and arcuate fasciculus (0.33-0.35 µm). The difference from the $R_2$ model was largest in the splenium, optic radiation, inferior fronto-occipital and inferior longitudinal fasciculi (0.12-0.19 µm) and smallest in the rostral body, anterior midbody, third branch of the superior longitudinal fasciculus (SLF III) and cingulum (0.02-0.06 µm).

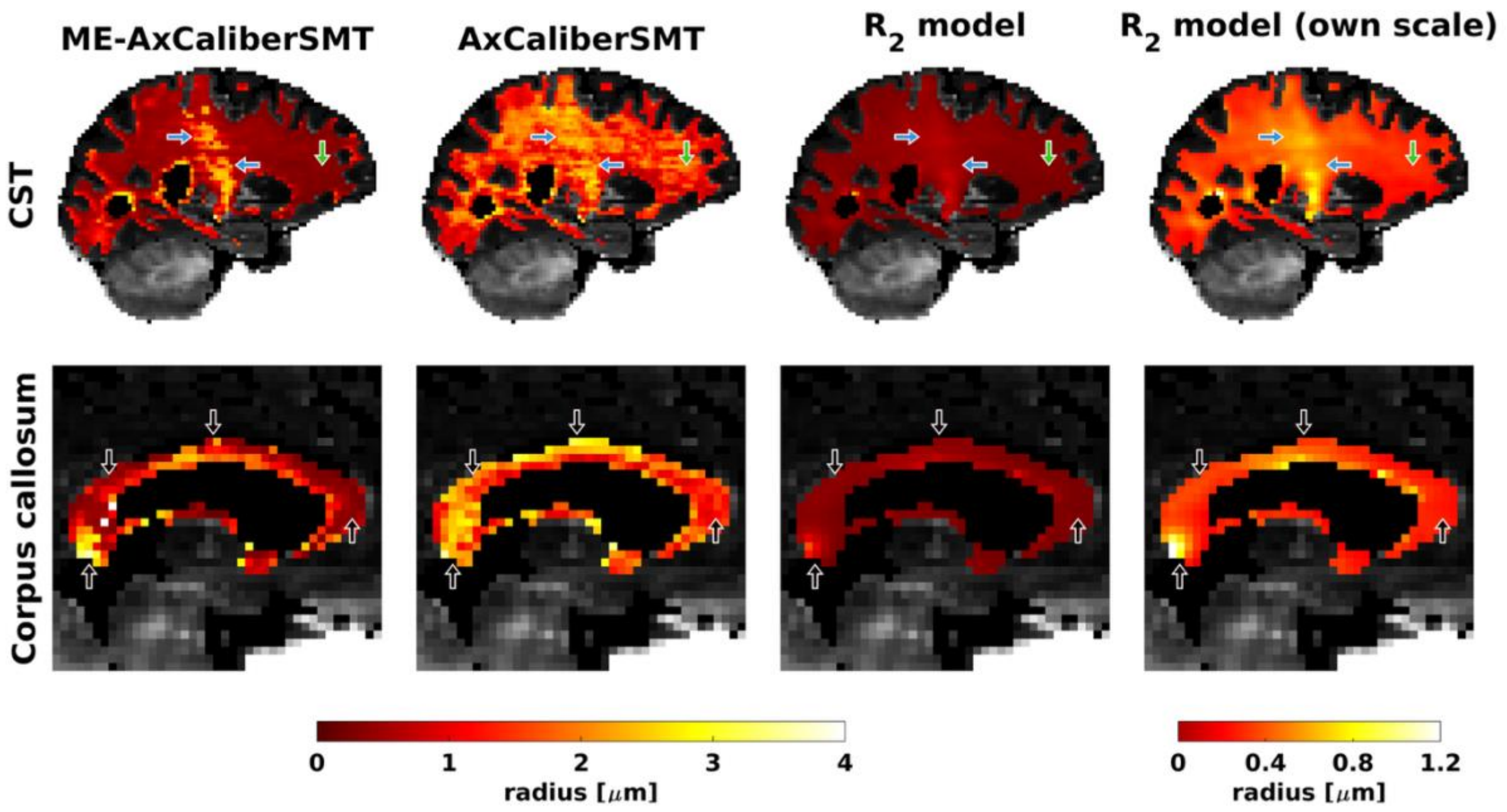

Figure 5: In vivo radius maps of one participant on Connectome 2.0. Sagittal slice through the corticospinal tract (CST; top); midsagittal slice through the corpus callosum (bottom). ME-AxCaliberSMT, AxCaliberSMT and the $R_2$ model (left to right) on a shared scale, and the $R_2$ model on its own scale. Effective radii (posterior medians for ME-AxCaliberSMT and AxCaliberSMT) are shown in white matter over the ME-AxCaliberSMT axonal fraction. The $R_2$ model was fitted to the dedicated multi-echo acquisition. Blue arrows highlight the CST; green arrows highlight the frontal white matter; black arrows highlight the low-high-low-high spatial contrast in CC from anterior to posterior segments.

Correspondence to Kwok-Shing Chan

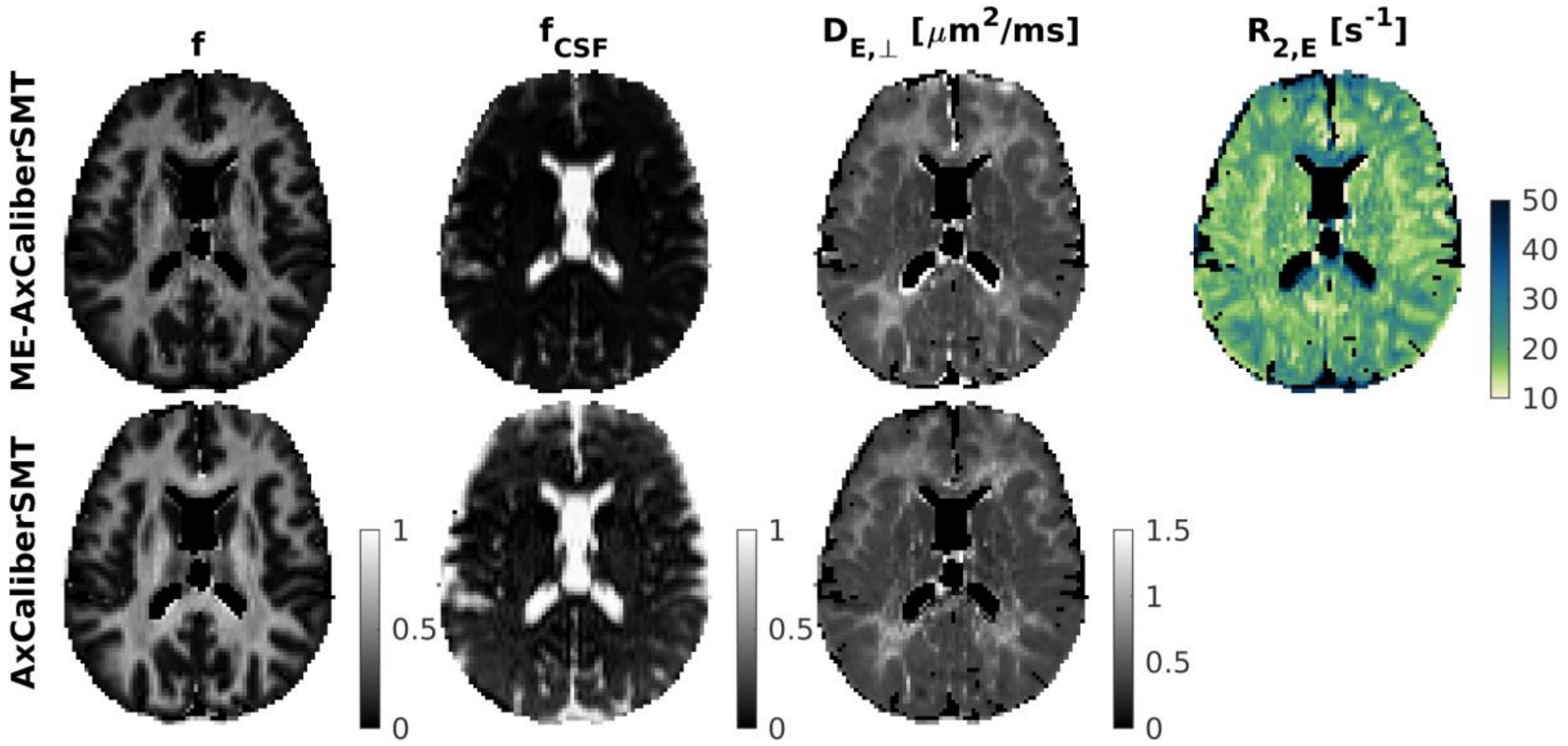

Figure 6: Example other fitted parameters in an axial slice on Connectome 2.0: axonal fraction $f$, CSF fraction $f_{CSF}$ and extra-axonal radial diffusivity $D_{E,\perp}$ for ME-AxCaliberSMT (top) and AxCaliberSMT (bottom), and the extra-axonal relaxation rate $R_{2,E}$, which only ME-AxCaliberSMT estimates. The signal fractions of AxCaliberSMT are $T_2$-weighted at TE = 51 ms, whereas those of ME-AxCaliberSMT are proton-density fractions.

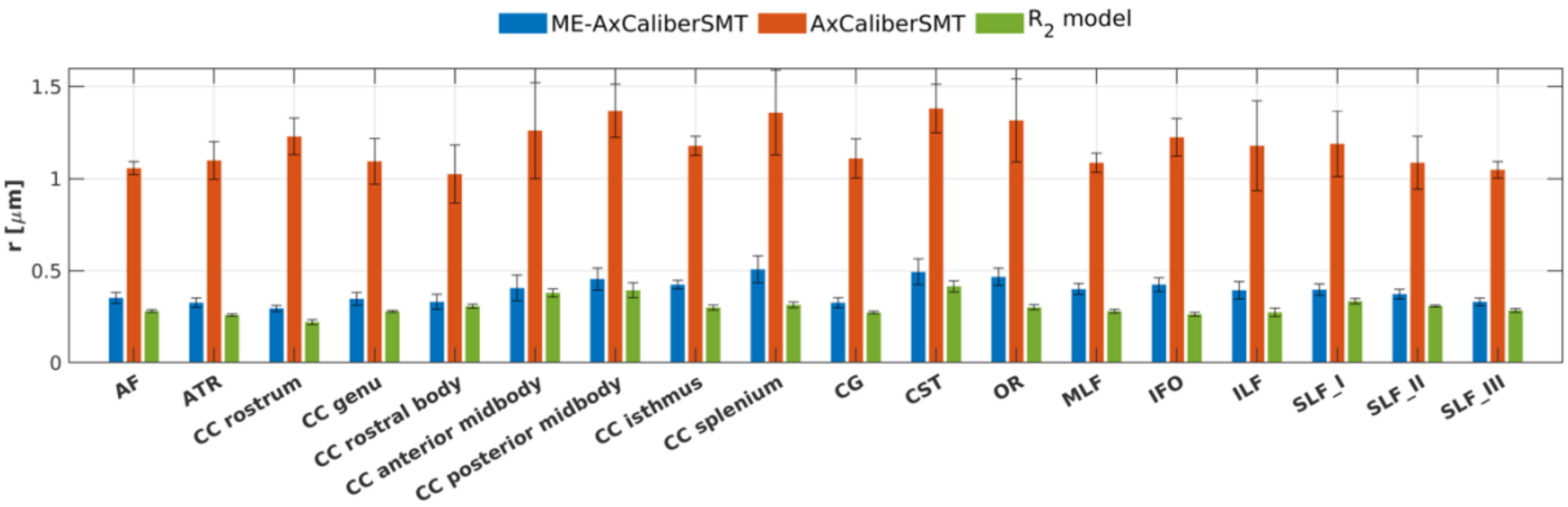

Figure 7: Bundle-wise axon radius on Connectome 2.0 for ME-AxCaliberSMT, AxCaliberSMT and the $R_2$ model. For each participant, the bundle value is the median radius within the TractSeg bundle; bars show the mean and error bars the standard deviation over the four participants. AF, arcuate fasciculus; ATR, anterior thalamic radiation; CC, corpus callosum (segments from anterior to posterior); CG, cingulum; CST, corticospinal tract; OR, optic radiation; MLF, middle longitudinal fascicle; IFO, inferior fronto-occipital fascicle; ILF, inferior longitudinal fascicle; SLF_I-III, superior longitudinal fascicle, branches I-III.

## 4.5 Agreement between scanners

The ME-AxCaliberSMT maps had the same pattern on both scanners, with larger radii along the CST and slightly higher values on Cima.X (Figure 8a). The AxCaliberSMT map was higher overall on Cima.X and largely lost the contrast between the corticospinal tract and the surrounding white matter.

Bundle-wise ME-AxCaliberSMT radii correlated between scanners (Pearson correlation 0.63-0.83 per participant, mean 0.73; Spearman 0.81), whereas AxCaliberSMT radii did not (-0.31 to -0.15, mean -0.23; Figure 8b). Both models gave higher values on Cima.X (Figure 8c). For ME-AxCaliberSMT, the bias was +0.21 μm (95% limits of agreement −0.07 to +0.50 μm; CCC = 0.24), with bundle values of 0.26-1.01 μm on Connectome 2.0 and 0.37-1.13 μm on Cima.X. For AxCaliberSMT, the bias was +0.63 μm (+0.02 to +1.25 μm; CCC = -0.03), with ranges of 0.85-2.00 and 1.59-2.35 μm. The cerebellar peduncles were an exception, with lower Cima.X values in two participants; without them, the mean ME-AxCaliberSMT correlation was 0.78. Bundle values of the two models correlated with a mean of 0.67 on Connectome 2.0 and 0.25 on Cima.X.

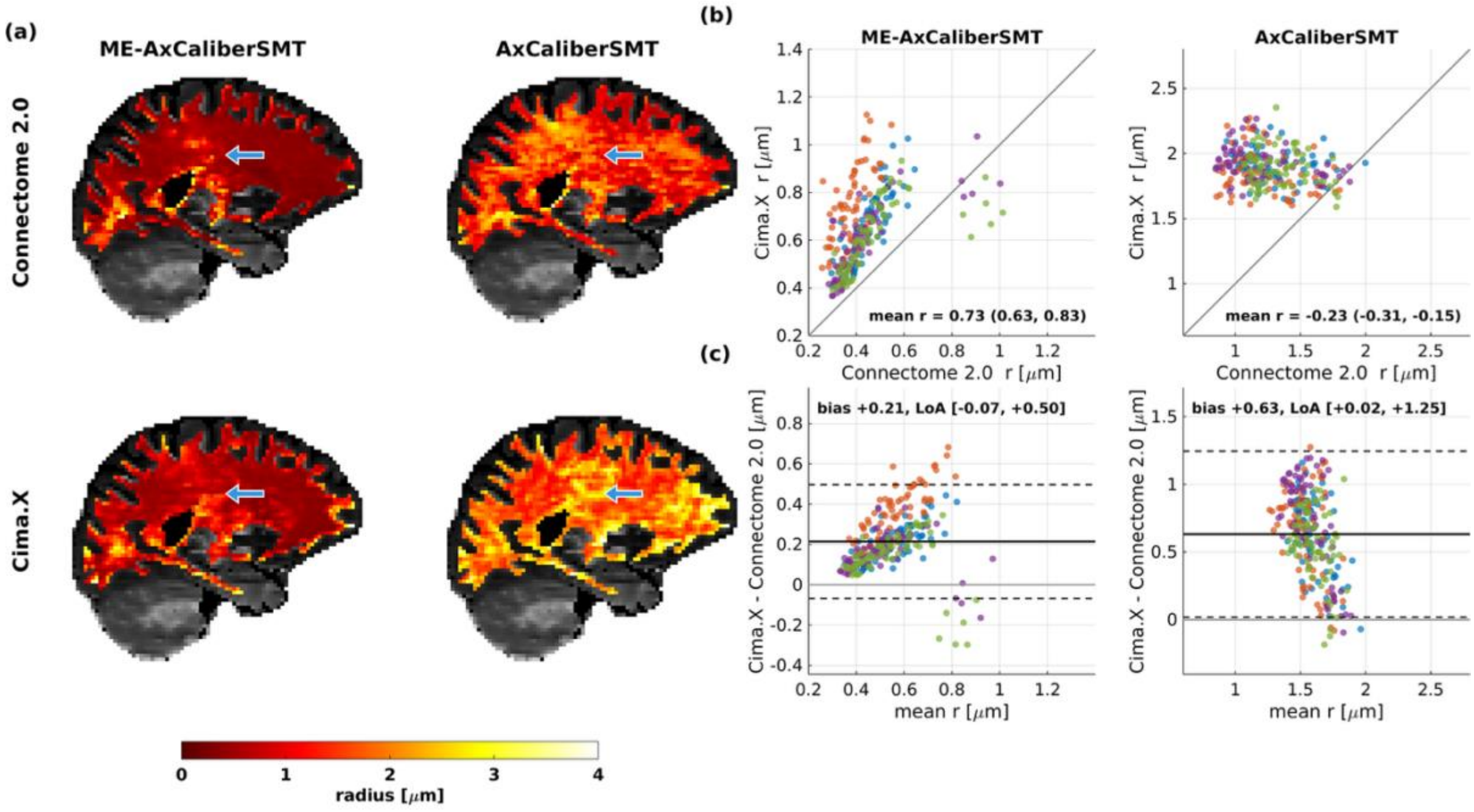

Figure 8: Agreement between Connectome 2.0 and Cima.X. (a) Effective radius maps in a sagittal slice through the corticospinal tract on both scanners for ME-AxCaliberSMT and AxCaliberSMT, on a shared scale. (b) Bundle-median radius on Cima.X against Connectome 2.0 for the four participants (colored) scanned on both systems (69 bundles × 4 participants = 276 bundle pairs; fornix excluded), with the identity line and the mean (range) over participants of the Pearson correlation across bundles. (c) Bland-Altman plots of the same data: solid line, mean difference (bias); dashed lines, 95% limits of agreement.

# 5. Discussion

Adding a second echo time, together with the radius dependence of intra-axonal relaxation, gives a second source of sensitivity to axon radius. In ME-AxCaliberSMT, small radii are determined mainly by relaxation and large radii by diffusion, with a sharp and predictable transition between the two. In simulations, this reduced the overall radius error 2.3-fold compared with AxCaliberSMT, and in vivo the model gave a bundle-wise pattern that was reproduced on a commercial 200 mT/m system, whereas AxCaliberSMT showed no consistent pattern.

## 5.1. Complementary sensitivity of diffusion and relaxation

The information carried by diffusion and relaxation differs by a factor of $(r/r^{*})^{10}$ (Eq. [8]), so the estimate switches from one mechanism to the other over a narrow range of radii (Eq. [12]). The transition radius increases with echo spacing and diffusion time and decreases with *b*, and was 1.42 μm on Connectome 2.0 and 1.85 μm on Cima.X, which had a shorter echo spacing (41 vs 28 ms) and weaker diffusion weighting. On Connectome 2.0, $r^{*}$ is close to the minimum radius detectable by diffusion alone, about 1.25 μm[23], so relaxation takes over where diffusion loses sensitivity. Below $r^{*}$, the diffusion signal changes too little to resolve radius[20–22], while relaxation is at its most sensitive, which is why ME-AxCaliberSMT recovered radii down to 0.1 μm in simulation, whereas AxCaliberSMT estimates below 1 μm carried little information on the true radius (Figures 2 and 4). Above $r^{*}$, ME-AxCaliberSMT converges to AxCaliberSMT, keeping the diffusion estimate where it is reliable, which also limits the effect of calibration errors (Figure 3d-e; Section 5.4). Compared with the $R_2$ model, it was more precise for large axons (SD of the error = 0.11 vs 0.68 μm above 2.7 μm) and accounted for the extra-axonal signal that remains at moderate *b*-values. Another advantage of ME-AxCaliberSMT is that it also measures the TE-independent proton density-weighted compartmental signal fractions, which can be useful to investigate oedema and axonal loss together with radius changes in neurological disorders and ageing[2,3,57].

Correspondence to Kwok-Shing Chan

## 5.2. Interpretation of the in vivo effective radius

With a distribution of radii, ME-AxCaliberSMT reports $w_D r_D + (1 - w_D) r_R$. Most white matter voxels in our data (82-94%) were below $r^*$, in the relaxation-dominated regime, where the reported effective radius is $r_R = \langle r^2 \rangle / \langle r \rangle = \langle r \rangle (1 + CoV^2)$. Any variations in radius distribution would therefore exceed the mean radius of the population (e.g., by 25-50% for CoV values of 0.5-0.7). The diffusion effective radius, $r_D = (\langle r^6 \rangle / \langle r^2 \rangle)^{1/4}$, is instead heavily dominated by the largest axons[13,24].

This difference in weighting explains much of the gap between the two models in vivo. The ME-AxCaliberSMT values fall within the range of histological inner radii in human white matter, where most axons have radii below 0.5 μm[8,9], though these post-mortem radii were not corrected for volume changes during fixation and embedding. A like-for-like comparison should also compute $r_R$ from the histological radius distribution instead of comparing with its mean. The AxCaliberSMT radius in the CST on our Connectome 2.0 cohort, 1.38 μm on average across participants, agrees with the diameter of 2.66 ± 0.54 μm reported in our previous study on the same scanner[23].

The regional pattern was anatomically plausible. In the CC, ME-AxCaliberSMT reproduced the low-high-low-high pattern of the histological fibre composition, in which thin fibres are densest in the genu and large fibres peak in the posterior midbody and increase again locally in the posterior pole of the splenium[9]. The CST also had larger radii than the association bundles, and the $R_2$ model did not reproduce the larger radius of the splenium.

The difference between ME-AxCaliberSMT and the $R_2$ model was smallest in bundles where small axons predominate, such as the body of the CC, the cingulum and SLF III, where the estimate is determined by relaxation and both models report a radius close to $r_R$. In the splenium, optic radiation and inferior fronto-occipital and longitudinal fasciculi, where the difference was largest, a broader distribution of radii or more large axons would increase $w_D$ and raise the ME-AxCaliberSMT estimate above $r_R$ (Eq. [12]), whereas the $R_2$ model is biased downwards because it ignores intra-axonal diffusion (Figure 4), which at $b$ = 6 ms/μm$^2$ attenuates the signal of axons with $r$ = 3 μm by 36%.

Because the relaxation contrast is anchored to the histological calibration of $R_2^C$ and $k_{2,A}$ [32], absolute radii in the relaxation-dominated regime inherit this calibration, including its choice of histological radius summary, so relative differences between bundles are the more robust measure.

## 5.3. Agreement between scanners

On Cima.X, where $r^*$ was 1.85 μm, AxCaliberSMT bundle radii (1.6-2.4 μm) did not correlate with those on Connectome 2.0, whereas ME-AxCaliberSMT preserved the bundle ranking (mean correlation 0.73). This suggests that radius contrast in typical white matter can be measured without ultra-high-performance gradients if relaxation is included.

ME-AxCaliberSMT values were nevertheless 0.21 μm higher on Cima.X (CCC = 0.24), so absolute values cannot be interchanged between the two protocols. The effective radius weighting does not explain this offset, because the larger $r^*$ on Cima.X gives more weight to the smaller relaxation effective radius. A more likely cause is estimation bias from the acquisition differences. With a shorter echo spacing, weaker diffusion weighting and lower SNR, the Cima.X posteriors are wider and skewed to the right by the lower bound at zero, which pushes the posterior median upwards. Population and spatial priors[54] should reduce this bias and are worth exploring in future cross-platform validation studies.

## 5.4. Calibration of the relaxation term

Errors in the fixed $R_2^C$ and $k_{2,A}$ propagate into $r$ in two ways: an error in $k_{2,A}$ rescales the relaxation effective radius, whereas an error in $R_2^C$ shifts $1/r$ by $\Delta R_2^C/k_{2,A}$ and affects large radii most. The $R_2$ model, which has no other source of information on $r$, was therefore the most sensitive to $R_2^C$, whereas diffusion bounded the error of ME-AxCaliberSMT at large radii (Figure 3).

Intra-axonal $R_2$ may vary between tracts, with fibre orientation relative to $B_0$, with age and with field strength, though the $B_0$ effect was shown to be small[31]. Estimating $R_2^C$ and $k_{2,A}$ together with $r$ from two echo times is poorly conditioned. The calibration also needs further validation in disease: white matter $T_2$ depends strongly on myelination in mouse models[58], but how demyelination affects the compartmental-specific axoplasm $R_2^C$ and $k_{2,A}$ remains unclear. Calibrating these constants against histology in more tracts, or with more echo times, remains an important next step.

## 5.5 Limitations

First, axons were modelled as impermeable, straight cylinders with a single intrinsic $R_2$ and no exchange, although calibre variations and beading[59–61] change both the surface-to-volume ratio and the diffusion signal. The model also attributes all restricted, stick-like signal to axons, yet glial processes can behave similarly at high *b*-values and confound diffusion-based axon diameter estimates[62]. The extra-axonal radial diffusivity was assumed to be independent of diffusion time[24], and the parallel diffusivities were fixed at 1.7 µm$^2$/ms. Second, the in vivo sample was small (five adults, four of whom were scanned on Cima.X on a different day), with no same-scanner scan-rescan data or histological validation, and the overlapping TractSeg bundles are not independent. Finally, the protocols took 66-68 min. Optimising the acquisition protocol is therefore needed for practicality. The two echo times are currently acquired in separately; acquiring both within the same excitation using advance acquisition methods, as in PRIME[63] or EPTI[64], could further shorten the scan time.

Correspondence to Kwok-Shing Chan

# 6. Conclusions

ME-AxCaliberSMT combines restricted diffusion and radius-dependent intra-axonal relaxation in one multi-echo spherical-mean model. Small radii are determined mainly by relaxation and large radii by diffusion, with a sharp, protocol-dependent transition at 1.4-1.9 μm, so the model reports the relaxation-weighted effective radius in most white matter. In simulations, it reduced the radius error 2.3-fold compared with AxCaliberSMT and was much less sensitive to the relaxation calibration than the $R_2$ model. In vivo, it gave radii within the histological range and a bundle-wise pattern that was reproduced on a commercial 200 mT/m system.

## Acknowledgements

The authors thank Dr. Derek K. Jones for helpful discussion on development of the methodology. KC is supported by NWO/ZonMw under the Rubicon award number 04520232330012. This study was support by the Office of the Director (OD) of the National Institutes of Health (NIH) in partnership with the National Institute of Dental and Craniofacial Research (NIDCR) under award number DP5OD031854, the OD of the NIH under award number S10OD032184, the National Institute of Neurological Disorders and Stroke (NINDS) of NIH under award numbers R01NS118187, the National Institute of Biomedical Imaging and Bioengineering (NIBIB) under award numbers P41EB015896, P41EB030006, and U01EB026996, the National Center for Research Resources (NCRR) of NIH under award numbers S10RR023401 and S10RR019307, and the National Institute on Aging under award number R21AG085795. BB is supported by NIH under award numbers R01EB032378, R01EB028797, UH3EB034875, R21AG082377 and R01EB034757.

## Data and Code Availability

The data and code that support the findings of this study will be made available upon reasonable request. The GPU processing tool for the ME-AxCaliberSMT model parameter estimation is available athttps://github.com/kschan0214/gacelle.

Correspondence to Kwok-Shing Chan

# Declaration of Interests

The authors declare that they have no known competing financial interests or personal relationships that could have appeared to influence the work reported in this paper.

Correspondence to Kwok-Shing Chan

# Author Contributions

**Kwok-Shing Chan**: Conceptualisation, Methodology, Software, Validation, Formal analysis, Investigation, Writing – Original Draft, Visualisation, Funding acquisition
**Yohan Jun**: Software, Methodology, Writing – Review & Editing
**Aneri Bhatt**: Data curation, Resources, Writing – Review & Editing
**Berkin Bilgic**: Methodology, Writing – Review & Editing
**Susie Y. Huang**: Methodology, Writing – Review & Editing, Supervision, Funding acquisition
**Hong-Hsi Lee**: Conceptualisation, Methodology, Software, Writing – Original Draft, Supervision, Funding acquisition

# Supporting Information

# S1. Effective radius of ME-AxCaliberSMT: derivation and numerical evaluation

**Linearised estimate:** With a distribution of radii, the diffusion and relaxation parts of the signal correspond to different effective radii, $r_D$ and $r_R$, whereas the ME-AxCaliberSMT model has a single effective radius $\hat{r}$. To find the radius that the fit returns, we write the signal as $S_{ME-AxcaliberSMT}(\theta, r_D, r_R)$, where $\theta = (f, f_{CSF}, D_{E,\perp}, R_{2,E})$ are the other fitted parameters and $r_D$ and $r_R$ enter only the diffusion and the relaxation terms, respectively. This decomposition rests on two assumptions. First, the mixture signal is separable into a diffusion term governed by a single effective radius $r_D$ and a relaxation term governed by a single effective radius $r_R$. For relaxation, this follows from the first-order expansion in $TE \cdot k_{2,A}/r$ that gives Eq. [10]. For diffusion, $r_D$ is defined numerically as the single radius whose cylinder signal best fits that of the mixture across all shells, so no assumption of small radial attenuation is needed. This radius has no closed form in general, because the sum of cylinder signals over the distribution cannot be reproduced exactly by a single radius at every b and Δ, so it depends on the protocol; Eq. [11] is its analytical limit to leading order in $b \cdot D_{A,\perp}$ and is used only for comparison. Separability neglects the cross-weighting between the two terms, which is small because the relaxation weighting varies little across radii ($TE \cdot k_{2,A} \approx$ 0.10 μm on Connectome 2.0 and 0.07 μm on Cima.X). Second, $r_D$, $r_R$ and the fitted $\hat{r}$ lie close enough to a common reference radius $r_0$ for the signal to be linear in each of them, and the noise of the spherical mean is Gaussian with equal variance in every shell, so that the posterior median within the bounds is close to the least-squares solution. Under these assumptions, the data and the single-effective-radius model are expanded to first order Taylor expansions around $r_0$ and the true nuisance parameters $\theta_0$, with $S_{ME-AxCaliberSMT,0} = S(\theta_0, r_0, r_0)$, the reference signal evaluated with $r_0$ in place of both $r_D$ and $r_R$:

$$y \approx S_0 + j_D\,(r_D - r_0) + j_R\,(r_R - r_0), \qquad S_{ME-AxcaliberSMT} \approx S_0 + J_\theta\,\Delta\theta + j\,\Delta r \quad Eq.[S1]$$

Correspondence to Kwok-Shing Chan

where $y$ is the signal of the voxel with a distribution of radii, $\Delta\theta = \theta - \theta_0$, $\Delta r = r - r_0$ and $j_D = \partial S/\partial r_D$, $j_R = \partial S/\partial r_R$, $j = j_D + j_R$ and $J_\theta = \partial S/\partial\theta$. With $P$ the projection orthogonal to the columns of $J_\theta$, the least-squares solution for $r$ is

$$\hat{r} = w_D\, r_D + (1 - w_D)\, r_R, \qquad w_D = \frac{(Pj)^{\mathrm{T}}(Pj_D)}{(Pj)^{\mathrm{T}}(Pj)} \; Eq. [12]$$

Because $j = j_D + j_R$, the weights of $r_D$ and $r_R$ sum to one. If the cross term $C = (Pj_D)^T(Pj_R)$ between the two sensitivities is small, $w_D = (I_D + C)/(I_D + I_R + 2C)$ reduces to $I_D/(I_D + I_R)$, the form used in Section 2.2. This holds when the projected diffusion and relaxation sensitivities are nearly uncorrelated across shells, which is favoured by acquiring every $b$-value at both echo times, so that radius changes through diffusion (mainly along $b$) and through relaxation (mainly along TE) are distinguishable after the nuisance parameters are projected out. The error of the simplified form, $C(I_R - I_D)/[(I_D + I_R)(I_D + I_R + 2C)]$, also vanishes at $r^*$, where $I_D = I_R$. Because $w_D$ depends on $r$, it was evaluated at the fitted radius by fixed-point iteration. Re-centring $r_0$ on $\hat{r}$ in this way keeps the linearisation error small; the error grows with the separation between $r_D$ and $r_R$, that is, with the width of the radius distribution, and its size in practice is given by the deviation of the fits from Eq. [12].

**Numerical evaluation:** The signal was the exact spherical mean (l = 0 term) of every shell of the protocol, with a cylinder signal for the intra-axonal compartment[1], normalised to the $b$ = 0 signal at the shortest TE. The tissue parameters were those of the effective-radius simulation (Section 3.1.3; $f$ = 0.7, $f_{CSF}$ = 0.05, $D_{E,\perp}$ = 0.51 µm$^2$/ms, $R_{2,E}$ = 25 s$^{-1}$), and $R_2^C$ and $k_{2,A}$ were as in Table 1. $j_D$, $j_R$ and $J_\theta$ were computed by central finite differences. $w_D(r)$ was evaluated on 200 logarithmically spaced radii between 0.1 and 4 µm, and $r^*$ and the radii at which $w_D$ = 0.1 and 0.9 were obtained by interpolation. For each distribution of radii in that simulation, $r_R$ was computed with Eq. [10], and $r_D$ was computed numerically as the single radius whose cylinder signal best fits the $r^2$-weighted diffusion signal of the mixture, with $f$, $f_{CSF}$, $D_{E,\perp}$ and $R_{2,E}$ free and the relaxation radius held at $r_R$ (Nelder-Mead minimisation of the residual sum of squares weighted by the number of volumes per shell, best of four starts). Because this $r_D$ is defined by a fit rather than by a closed form, it is referred to below as the numerical $r_D$. Starting from $r_0$ = $r_R$, $r_0$ was updated to the mean of $r_0$ and the predicted $\hat{r}$ until the change was below $10^{-5}$ µm. For the $R_2$ model, which ignores intra-axonal diffusion, $r_R$ was also computed with each radius weighted by its spherical-mean diffusion attenuation $A(r_i)$ at $b$ = 6 ms/µm$^2$ ($\Delta/\delta$ = 13/6 ms), that is, $\sum r_i^2 A(r_i)/\sum r_i A(r_i)$. For the curves of Figure S1, $r_R$, the numerical $r_D$ and the Eq. [12] prediction were also computed on a dense grid of mean radii (0.1-2.2 µm in steps of 0.025 µm) for Gamma distributions with CoV($r$) = 0.25 and 0.5, as in that simulation, with each distribution integrated on a 0.005-µm radius grid instead of sampled.

**Results:** For the Connectome 2.0 protocol, $w_D$ was 0.06 at $r$ = 1 µm, 0.61 at 1.5 µm and 0.94 at 2 µm, giving $r^*$ = 1.42 µm and a 10–90% range of 1.08-1.88 µm (Figure 1d). For the Cima.X protocol, $w_D$ was 0.01, 0.16 and 0.64 at the same radii, giving $r^*$ = 1.85 µm and a range of 1.39-2.48 µm. Eq. [9], evaluated for the strongest diffusion weighting of each protocol ($b$ = 17.8 ms/µm$^2$, $\Delta/\delta$ = 30/6 ms, ΔTE = 41 ms on Connectome 2.0; $b$ = 6 ms/µm$^2$, $\Delta/\delta$ = 35/15 ms, ΔTE = 28 ms on Cima.X), gave 1.21 and 1.70 µm, respectively; the full calculation is larger because the other shells and the fitted nuisance parameters reduce the diffusion information. The slope of logit($w_D$) against log $r$ was 7.8 for Connectome 2.0, compared with 10 for the scaling $(r/r^*)^{10}$ of Section 2.2. Across the 16 conditions of the effective-radius simulation with CoV($r$) > 0, the ME-AxCaliberSMT fits deviated from the prediction of Eq. [12] by 0.022 µm (RMS), compared with 0.191 µm when $r_D$ was taken from Eq. [11], 0.300 µm from $r_R$ alone, 0.123 µm from the numerical $r_D$ alone and 0.236 µm from $r_D$ of Eq. [11] alone (Figure S1). The remaining offset, 0.01-0.05 µm above the prediction for small mean radii and broad distributions, is consistent with the first-order expansion in $TE \cdot k_{2,A}/r$ that underlies $r_R$. For the $R_2$ model, weighting by the attenuation at $b$ = 6 ms/µm$^2$ lowered $r_R$ at CoV = 0.5 from 1.25 to 1.22 µm at $\langle r \rangle$ = 1 µm, from 1.88 to 1.72 µm at 1.5 µm and from 2.49 to 2.14 µm at 2 µm.

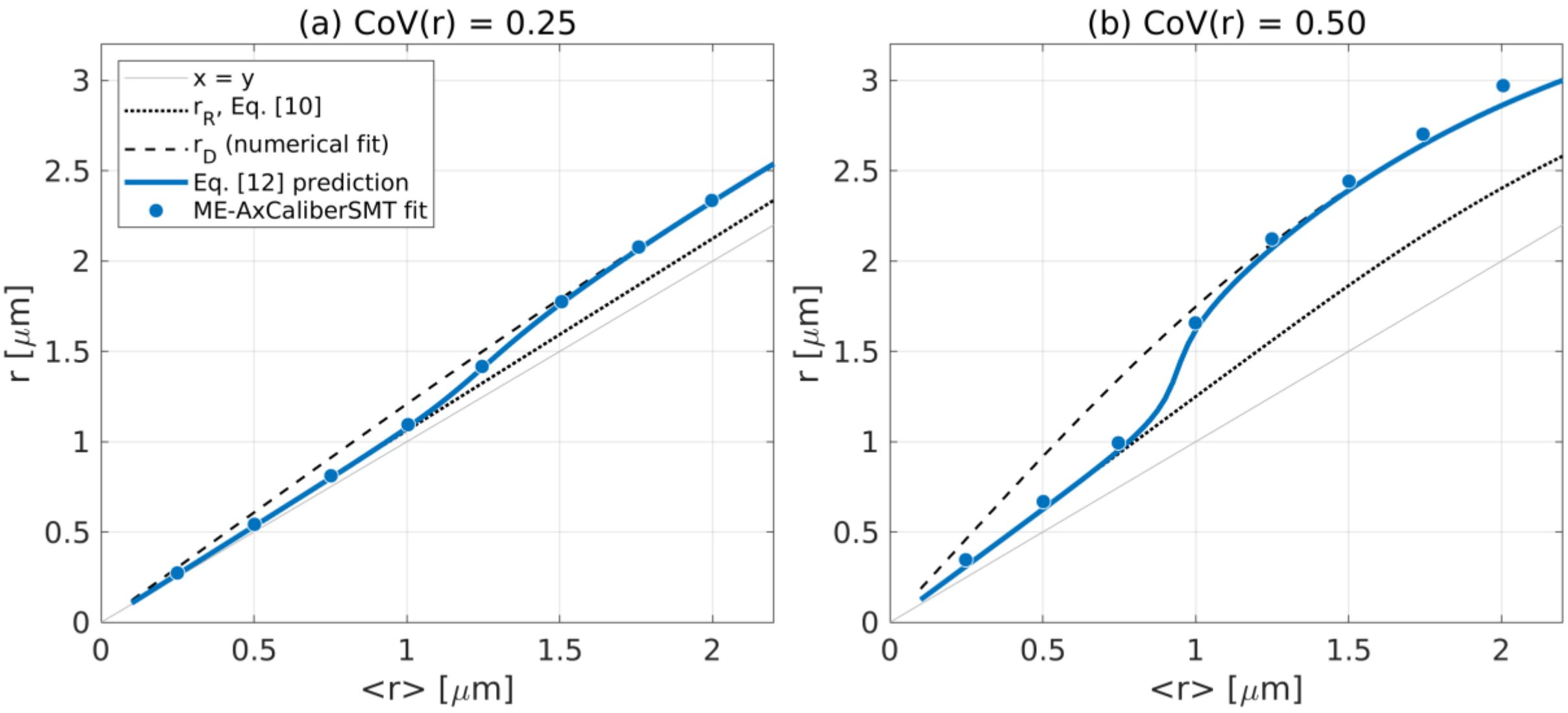

**Figure S1.** Weighting of the diffusion and relaxation effective radii in the effective-radius simulation (Section 3.1.3), for CoV($r$) = 0.25 (a) and 0.5 (b): ME-AxCaliberSMT fits (median over 50 voxels of the posterior median) at the eight simulated mean radii, with the prediction of Eq. [12], the relaxation effective radius $r_R$ (Eq. [10]) and the numerical diffusion effective radius $r_D$.

Correspondence to Kwok-Shing Chan

## Supplementary Reference

Correspondence to Kwok-Shing Chan